\documentclass[fleqn,twoside]{article}
\usepackage{espcrc2}
\usepackage{epsfig}

\newcommand{\AmS}{{\protect\the\textfont2
  A\kern-.1667em\lower.5ex\hbox{M}\kern-.125emS}}

\newcommand{\ba}{\begin{eqnarray}}
\newcommand{\ea}{\end{eqnarray}}
\newcommand{\bas}{\begin{eqnarray*}}
\newcommand{\eas}{\end{eqnarray*}}
\newcommand{\be}{\begin{equation}}
\newcommand{\ee}{\end{equation}}
\newcommand{\bes}{\begin{equation*}}
\newcommand{\ees}{\end{equation*}}
\newcommand{\bi}{\begin{itemize}}
\newcommand{\ei}{\end{itemize}}

\newcommand{\bcentre}{\begin{center}}
\newcommand{\ecentre}{\end{center}}

\font\tenmsb=msbm10 scaled\magstep1
\font\sevenmsb=msbm7 scaled\magstep1
\font\fivemsb=msbm5 scaled\magstep1
\newfam\msbfam
\textfont\msbfam=\tenmsb
\scriptfont\msbfam=\sevenmsb
\scriptscriptfont\msbfam=\fivemsb

\newcommand{\order}[1]{{\mathcal O}(#1)}

\newcommand{\ksea}{\kappa^{\rm sea}}
\newcommand{\kval}{\kappa^{\rm val}}

\title{
\hfill\begin{minipage}{0pt}\scriptsize \begin{tabbing}
	\hspace*{\fill} Edinburgh-2002/12\\ \end{tabbing}\end{minipage}\\[8pt]
	\vspace{-1cm}
Excitations of the nucleon with dynamical fermions}

\author{UKQCD collaboration, C.M.~Maynard\address[ed]{School of Physics,
	JCMB, Kings Buildings, University of Edinburgh, Edinburgh, EH9
	3JZ, UK}\\ 
	LHP collaboration, D.G.~Richards\address{Jefferson
	Laboratory, MS 12H2, 12000 Jefferson Avenue, Newport News, VA
	23606, USA}}

\begin{document}

\begin{abstract}
We measure the spectrum of low-lying nucleon resonances using Bayesian
fitting methods.  We compare the masses obtained in the quenched
approximation to those obtained with two flavours of dynamical
fermions at a matched lattice spacing.  At the pion masses employed in
our simulations, we find that the mass of the first positive-parity
nucleon excitation is always greater than that of the parity partner
of the nucleon.
\vspace{1pc}
\end{abstract}

\maketitle

The observed spectrum of excited nucleon states contains some puzzles,
such as the anomalously light masses of the Roper $N^*(1440)$ and
$\Lambda(1405)$. Neither of these states naturally lies within the
quark model, and this had led to speculation as to whether they are
really three quark states.  This spectrum, at least in principle, is
accessible to lattice QCD calculations. In this study we
extract the masses of the nucleon ($N^{+}$), its parity partner
($N^-$) and the lightest positive-parity excitation ($N^{\prime +}$)
and examine their ordering.

Lattice determinations of the nucleon mass generally use the following three
interpolation operators:
\begin{eqnarray}
N_1^{1/2+} & = & \epsilon_{ijk} (u_i^T C \gamma_5 d_j) u_k\label{eq:N1},\\
N_2^{1/2+} & = & \epsilon_{ijk} (u_i^T C d_j) \gamma_5 u_k\label{eq:N2},\\
N_3^{1/2+} & = & \epsilon_{ijk} (u_i^T C \gamma_4 \gamma_5 d_j) u_k.
\label{eq:N3}
\end{eqnarray}
These operators have an overlap with both positive- and negative-parity states,
but on a lattice with (anti-)periodic boundary conditions we can use the
parity projection operator $\frac{1}{2}(1\pm\gamma_4)$ to delineate
forward and backward propagating states of opposite parities. The time
dependence of the parity projected correlators is then
\be
\label{eqn:sum_of_exps}
  C_{\pm}(t)=\sum_n A_{n}^{\pm} e^{-M_{n}^{\pm} t} + A_{n}^{\mp}
e^{-M_{n}^{\mp} (N_t-t)}.
\ee

The ``diquark'' part of $N_{1,3}$ couples the upper spinor components,
whilst the $N_2$ operator couples upper and lower components, and so
vanishes in the non-relativistic limit. The expectation, confirmed by
the data, is that the $N_{1,3}$ operators give a much better overlap
with the nucleon ground state while the $N_2$ operator has an overlap
with the lowest positive-parity nucleon excitation.

\begin{figure}
\epsfig{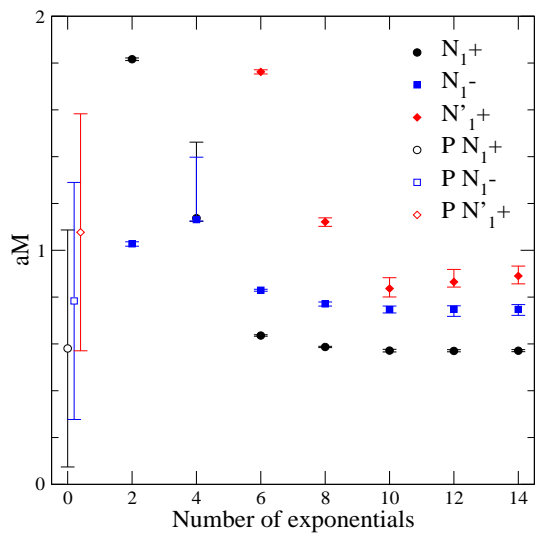}
\vspace{-1.0cm}
\caption{Mass versus number of exponentials. $\beta=6.2$, $\kappa=0.1346$.}
\label{fig:nexp}
\vspace{-0.75cm}
\end{figure}

\begin{figure}
\epsfig{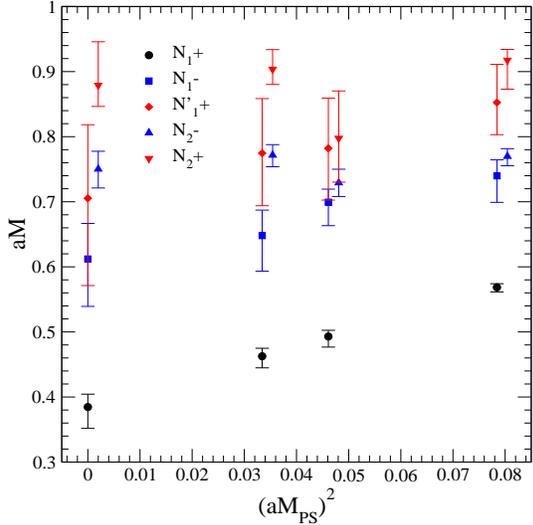}
\vspace{-1.0cm}
\caption{Quark mass dependence of baryon masses at $\beta=6.2$}
\label{fig:B62_chi}
\vspace{-0.75cm}
\end{figure}

For the negative-parity states the signal is short lived
\cite{neg_par_baryons}, so the choice of fit range may lead to a substantial
systematic uncertainty.  The aim of Bayesian fitting is to eliminate this
uncertainty~\cite{gpl_bayes}.  Furthermore, by fitting all
the data, it may be possible to measure the contribution from excited
states.  The number of exponentials $N_{\rm exp}$ is chosen such that
the results for the ground and first excited states are stable if
$N_{\rm exp}$ is increased. The ground-state priors were guessed from
the correlator data at large temporal separations using the effective
mass, 
\be 
M^{\rm eff}(t)=\ln\left(\frac{C(t+1)}{C(t)}\right), 
\ee
and priors for the mass splitting and excited state amplitudes
obtained using $ \Delta M = M_1^{\rm eff}(\tau) - M_0^{\rm eff}(t)$,
where $\tau < t$, and $M_1^{\rm eff}$ is determined from the
correlator $C^\prime$ with the ground state prior contribution
subtracted.  This ensures that the minimisation algorithm starts in
the neighbourhood of the global minimum of $\chi^2$.  The width of the
priors is chosen to be larger than the splitting.  Furthermore, the
fits are more stable if both $M_0$ and $\Delta M$ are constrained to
be positive, so instead of fitting to eqn (\ref{eqn:sum_of_exps}) we
fit the following expression to the data \be f(t)=\sum_n
A_n\exp\left[-\sum_{m=0}^ne^{\lambda_m}t \right] \ee where
$\lambda_k=\ln\mu_k$, $\mu_0=m_0$ and $\mu_k=M_k-M_{k-1}$.

The calculation was performed using the standard Wilson gluon action,
and the non-perturbatively $\order{a}$-improved
Sheikholeslami-Wohlert action. Lattice
artefacts thus appear in the spectrum at $\order{a^2}$.  The values of
$\beta$ and $\ksea$ for the coarsest quenched (Q) and the
dynamical-fermion (DF) ensembles were chosen so that the lattice
spacing is the same. The light hadron spectrum has already been
determined by the UKQCD collaboration for $\beta=6.2$ and
$\beta=6.0$~\cite{QLHS} and for DF matched ensembles~\cite{DFLHS}. The
$N^-$ masses have also been determined 
for $\beta=6.2$ and $\beta=6.0$~\cite{neg_par_baryons}.

\begin{table}
\caption{Details of the data sets. $a^{-1}$ in GeV, the $m_\pi/m_\rho$ values
for the DF configurations are at $\ksea=\kval$.}
\label{tab:data}
\begin{tabular}{ccccc}
\hline
$\beta$,$\ksea$& cfgs& $a^{-1}$& V& $m_\pi/m_\rho$ \\\hline
$6.2$,$0$ &     $216 $ & $2.91$ & $24^3\times48$&N/A \\
$6.0$,$0$ &     $496 $ & $2.12$ & $16^3\times48$&N/A \\
$5.93$,$0$ &$623 $ &$1.90$ & $16^3\times32$&N/A	 \\\hline
$5.2$,$0.1350$ &$202 $ &$1.91$ & $16^3\times32$&$0.70(1)$  \\
$5.26$,$0.1345$ &$102 $ &$1.90$ & $16^3\times32$&$0.78(1)$  \\
$5.29$,$0.1340$ &$101 $ &$1.94$ & $16^3\times32$&$0.84(1)$  \\
\hline
\end{tabular}
\vspace{-1.0cm}
\end{table}

We perform simultaneous fits to both the local-local and smeared-local
correlators of both parities.  However, for the lightest two DF
ensembles the limited number of configurations only allows a fit to
the local-local correlator. Thus we only fit to one correlator for all the DF 
ensembles. The dependence of
the extracted masses on the number of exponentials $N_{\rm exp}$ is
shown in figure \ref{fig:nexp}, together with the priors, shown as the
open symbols.  The results are stable when $N_{\rm exp} > 10$,
corresponding to five states of each parity.  For $\beta=\{6.2,6.0\}$
we use $N_{\rm exp} = 12$, and for the DF and matched Q ensembles we
use $N_{\rm exp} = 10$.

\begin{figure}
\epsfig{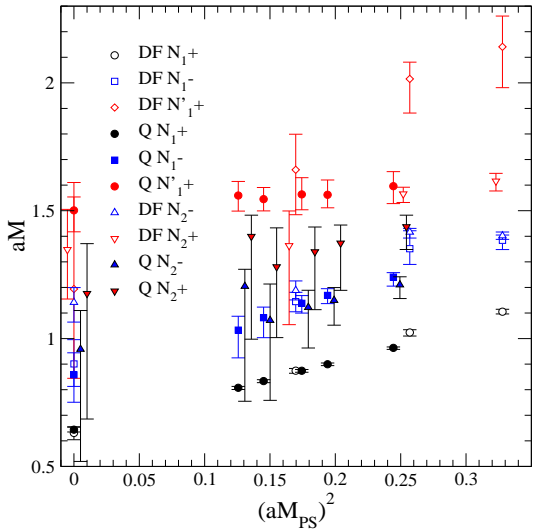}
\vspace{-1.0cm}
\caption{$N_f=2$ (open symbols) versus $N_f=0$ (closed symbols). Some
of the symbols have been offset for clarity.}
\label{fig:NF2_vs_NF0}
\vspace{-0.75cm}
\end{figure}

The spectrum as determined from the operators $N_{1,3}$ ($N_1$
hereafter) and $N_2$ is shown in figure~\ref{fig:B62_chi}; we find
consistency between the masses of $N^{-}$ using the two operators, and
between the excited-state mass $N^{\prime +}_1$ and ground-state mass
$N_2^+$.  However, as the quark mass is decreased the data quickly
become very noisy, and at lighter quark masses, the masses from the
$N_2$ operator are larger than those from the $N_1$. Only a simple
linear chiral extrapolation is attempted as the data is too noisy for
a more sophisticated approach.

\begin{figure}[t]
\epsfig{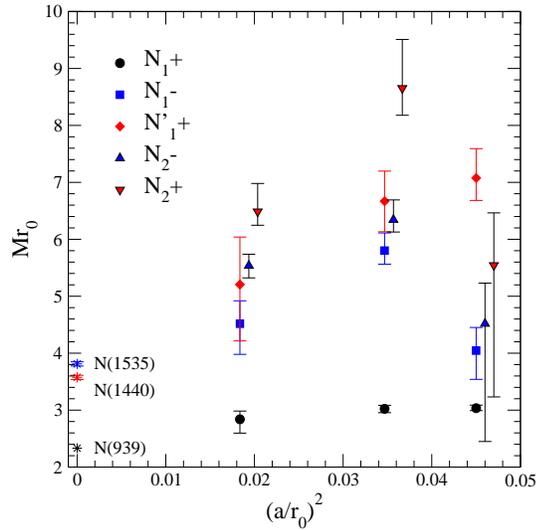}
\vspace{-1.0cm}
\caption{Lattice spacing dependence in quenched approximation to QCD.}
\label{fig:contin}
\vspace{-0.75cm}
\end{figure}

The comparison between the Q and DF ensembles at matched lattice
spacing is shown in figure \ref{fig:NF2_vs_NF0}, revealing no
systematic difference between the Q and DF results.  In all cases the
mass of $N^{\prime +}$ is above that of $N^-$.  Furthermore the
positive-parity excited state mass obtained from $N_1$ is consistent
with the ground-state mass obtained from $N_2$, except at light quark
masses where the latter becomes very noisy.  Finally, the lattice
spacing dependence of the measured masses on the quenched ensemble is
shown in figure \ref{fig:contin}.

We have measured the nucleon spectrum and found that the masses are
ordered according to $N^{\prime +} > N^- > N^+$ for both the Q and DF
data at all measured quark masses and lattice spacings, thus providing
little evidence for the Roper resonance as a simple three-quark
nucleon excitation.  However, there are several caveats. The
statistical quality of the data is poor.  The spatial extent is only
around $1.7~{\rm fm}$, though a comparison with a calculation of both
the $N^+$ and $N^-$ masses on a lattice of around twice that volume at
$\beta=6.0$ suggests that at these quark masses the finite-volume
corrections are small~\cite{neg_par_baryons}; such a comparison has
not been performed for $N^{\prime +}$.  Finally, the quark masses are
still large, with the smallest DF $m_{PS}/m_V = 0.7$. A calculation in
the quenched approximation to QCD using overlap fermions with
pseudoscalar masses as light as 200 MeV indicates a dramatic
cross-over between $N^-$ and $N^{\prime +}$ at light quark
masses~\cite{Lee_boston}.

The experimental study of the nucleon resonances is a crucial tool for
investigating the nature of the QCD interaction.  Lattice calculations
have a vital role both in interpreting the data, and in guiding future
experimental searches.

CMM acknowledges grants PPARC PPA/P/S/1998/00255, PPA/GS/1997/00655,
EU HPRN-CT-2000-00145-Hadrons/LatticeQCD,  and thanks C.T.H.~Davies
for useful discussions. This work was supported in part by DOE
contract DE-AC05-84ER40150 under which the Southeastern Universities
Research Association (SURA) operates the Thomas Jefferson National
Accelerator Facility.


\end{document}